\documentclass[letterpaper]{jpsj3}
\usepackage{txfonts}
\usepackage{graphicx}

\title{High-Field Torque Magnetometry on the Kagome Antiferromagnet Karpenkoite}

\author{
Hibiki Kunisawa$^1$,
Ryuya Watanabe$^1$,
Jun-ichi Yamaura$^2$, 
Yoshimitsu Kohama$^2$
and Toshihiro Nomura$^1$\thanks{nomura.toshihiro@shizuoka.ac.jp}
}
\inst{
$^1$Department of Physics, Shizuoka University, Shizuoka 422-8529, Japan \\
$^2$Institute for Solid State Physics, University of Tokyo, Kashiwa, Chiba 277-8581, Japan
}

\abst{
We report torque magnetometry results on single crystals of karpenkoite Co$_3$(V$_2$O$_7$)(OH)$_2$$\cdot$2H$_2$O, a model candidate for the kagome antiferromagnet.
No field-induced phase transition is detected up to 45~T for $B \parallel c$ and $B \parallel a$. 
Instead, the torque reveals a continuous spin reorientation toward saturation, most likely governed by a dominant Dzyaloshinskii–Moriya interaction.
}

\begin{document}
\maketitle

Geometrical frustration, where the lattice geometry prevents all pairwise interactions from being simultaneously satisfied, plays a key role in various emergent magnetic phenomena \cite{A.P.Ramirez}.
The kagome antiferromagnet (KAFM), composed of corner-sharing triangles, represents a prototypical case.
Antiferromagnetic coupling on this geometry strongly suppresses conventional long-range order and enhances the possibility of exotic quantum states such as spin liquids or valence-bond crystals \cite{Balents,Zhou}. 
The magnetization process of the spin $S=\frac{1}{2}$ KAFM is predicted to show plateaus at fractional values of the saturated moment (1/9, 1/3, 5/9, and 7/9) \cite{Hida,Nishimoto,Nakano,Capponi}, reflecting magnon crystallization and field-induced quantum spin-liquid phases.
Experimentally, magnetization plateaus have been observed in several KAFMs \cite{Okuma19,Okuma20,Jeon,Suetsugu,Kato,Yoshida}.
However, the high exchange coupling constants make it challenging to access the full magnetization processes, and especially the 5/9 and 7/9 plateaus have not been experimentally resolved.

In this study, we focus on the Co$^{2+}$-based perfect kagome antiferromagnet karpenkoite, Co$_3$(V$_2$O$_7$)(OH)$_2$$\cdot$2H$_2$O \cite{Kasatkin,Haraguchi}.
The spin orbit coupling and crystal-field effects lift the degeneracy of the 3$d$ orbitals, yielding a pseudospin $J_\mathrm{eff}=\frac{1}{2}$ kagome lattice.
Haraguchi \textit{et al.} reported the magnetization curve of powdered karpenkoite up to the saturation magnetic field ($\sim$30~T) \cite{Haraguchi}.
However, the magnetization curve shows no field-induced plateau, the hallmark of the KAFM.
This may be because the measured magnetization averages over the randomly oriented samples with an anisotropic $g$ factor.
In this paper, we report a method to grow single crystals of karpenkoite ($\sim$100~$\mu$m) and results of torque magnetometry on the single crystals up to the saturation magnetic fields.

Single crystals of karpenkoite were synthesized under hydrothermal conditions.
We used 3.5~g of Co(NO$_3)_2\cdot$6H$_2$O, 1.25~g of V$_2$O$_5$, and 1.2~g of NaOH as initial chemical reagents, all from Fujifilm Wako. 
The reagents and 22~ml of purified water were placed in a Teflon-lined stainless-steel autoclave and heated at 120${}^\circ$C.
After 3--30 days, orange single crystals with a regular hexagonal shape were obtained.
We picked up several transparent single crystals without apparent stacking faults along the $c$ axis.
We confirmed that the measured crystal was a single crystal of karpenkoite by using single-crystal x-ray diffraction (XRD).

\begin{figure}
\centering
\includegraphics[width=0.79\columnwidth]{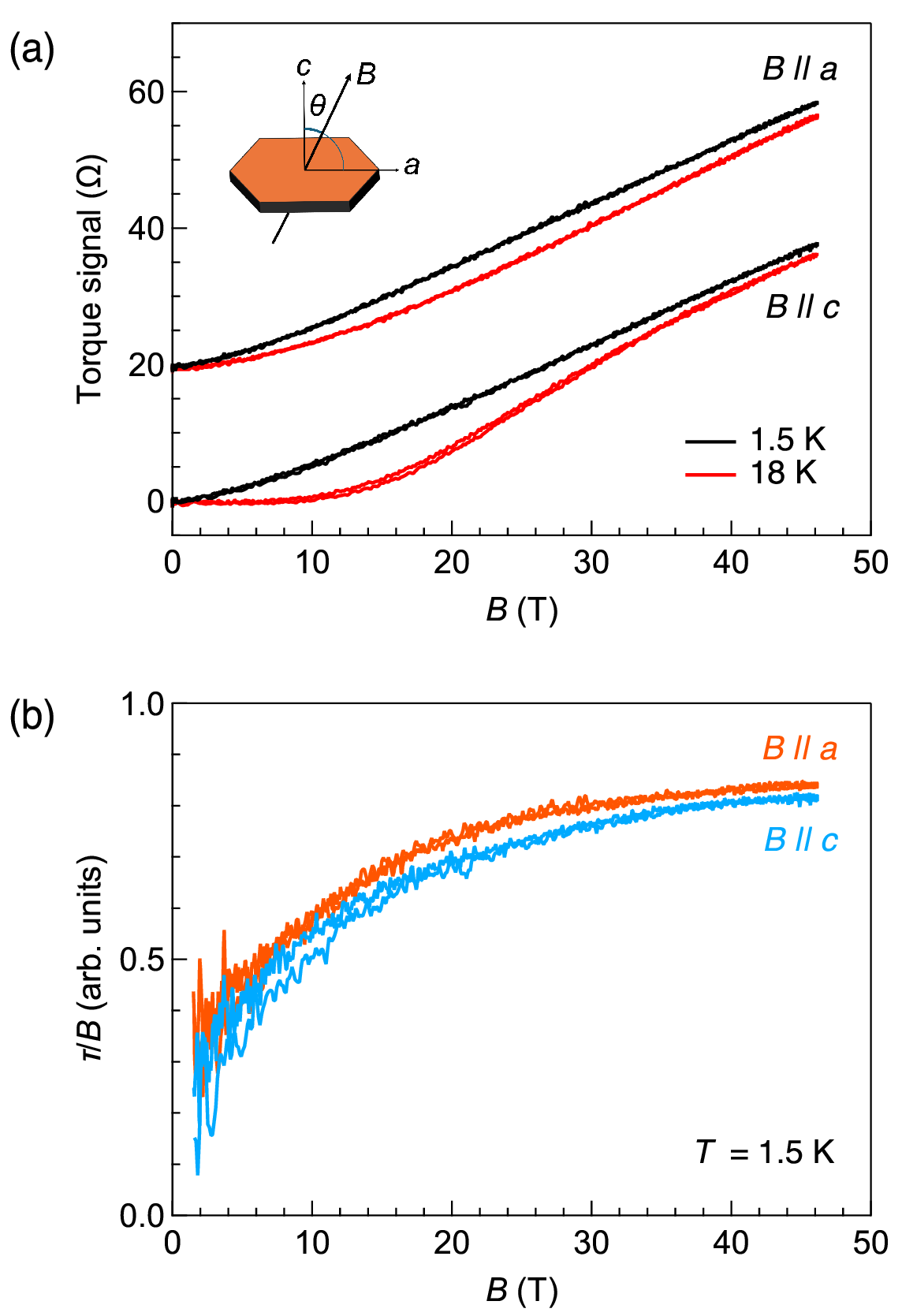}
\caption{
(a) Torque signal as a function of magnetic field nearly along the $c$ and $a$ axes. The black and red curves show the results at 1.5~K and 18~K, respectively. The curves for $B \parallel a$ are shifted for clarity.
The inset shows the definition of $\theta$.
(b) Torque signal normalized by the magnetic field ($\tau / B$) at 1.5~K.
}
\label{f1}
\end{figure}

High-field torque magnetometry was performed using a piezoresistive microcantilever (SEIKO PRC-120) in pulsed magnetic fields up to 45~T at the Institute for Solid State Physics, UTokyo, which is widely used to study fine structures of magnetization curves \cite{Ohmichi,Sebastian,Zheng}.
This technique can apply to a tiny crystal grown in this study and detects a torque signal proportional to ${\bf M} \times {\bf B}$.
A karpenkoite crystal ($\sim$30$\times$30$\times$2~$\mu$m$^3$) was mounted on the cantilever with Apiezon N grease, which also prevented dehydration in vacuum.
The sample stage was rotated within the $a$–$c$ plane, and the cantilever detected the torque signal in this plane.
The field orientation relative to the $c$ axis was $\theta = 1.9^\circ$ for the $B \parallel c$ and $\theta = 84.3^\circ$ for the $B \parallel a$ geometries (see also the inset of Fig.~\ref{f1}(a)).
The slight misalignment enhanced the torque signal under magnetic fields.
The cantilever resistance was measured using a digital lock-in technique at 50~kHz.
The torque signal was defined as the difference in resistance between the sample-mounted and reference cantilevers.

Figure~\ref{f1}(a) shows the torque signals for the field nearly along the $c$ and $a$ axes up to 45~T at 1.5 and 18~K.
The reported ordering temperature of karpenkoite was $T_\mathrm{N}=4.6$–4.9~K \cite{Haraguchi}; thus, the difference between the 1.5~K and 18~K data reflects the torque signal related to the arrangement of the ordered moments.
At both temperatures, the torque increases monotonically with field, indicating the absence of any magnetization plateau.
We note that the anomaly at 75 mT, which is related to weak ferromagnetism \cite{Haraguchi}, was not detected in our setup due to its lower sensitivity near zero magnetic fields.

We also performed torque magnetometry at 100~K, far above the ordering temperature.
The torque signal at 100~K, which we attribute it to the anisotropic $g$ factor caused by the crystal field, is approximately 20\% of the results at 1.5~K.
Hence, the dominant torque signal at 1.5~K originates from the ordered or short-range-ordered moments.
We note that the sign of the torque indicates easy-plane anisotropy over the entire temperature range.

Figure~\ref{f1}(b) displays the field-normalized torque, $\tau/B$, at 1.5~K.
Since $\tau \propto {\bf M} \times {\bf B}$, $\tau/B$ scales with the magnetization as long as the anisotropy does not change largely.
$\tau /B$ shows a smooth increase for both $B \parallel c$ and $B \parallel a$ geometries, similar to the magnetization curve for the powder sample \cite{Haraguchi}.
These results again indicate no field-induced phase transition.

We now discuss the origin of the missing plateau.
In the kagome lattice, the absence of inversion symmetry at the midpoint between Co ions allows for a sizable Dzyaloshinskii–Moriya (DM) interaction.
In karpenkoite, the DM strength is estimated as $|D|\!\sim\!10$~K, significantly larger than the nearest-neighbor exchange coupling $J\!\sim\!0.6(3)$~K \cite{Haraguchi}.
Although the latter may be underestimated considering the Weiss temperature of $\theta_\mathrm{W} = -17.4$~K \cite{Haraguchi,comment1}, the strong $D/J$ ratio indicates that the magnetism is dominated by DM interaction.
This large $D$ induces weak ferromagnetism \cite{Elhajal,Cepas} and suppresses the plateau state, leading to a finite slope of magnetization curve \cite{Yoshida}.

Magnetic anisotropy of the Co$^{2+}$ ions may also play a role.
The easy-plane anisotropy revealed by torque indicates that the system deviates from the ideal Heisenberg limit toward an XXZ model.
However, numerical studies suggest that most KAFM plateaus survive with anisotropy terms \cite{Kshetrimayum}.
Therefore, the in-plane anisotropy in karpenkoite cannot be the origin for the absence of the plateau.

In conclusion, we performed high-field torque magnetometry on single-crystalline karpenkoite up to 45~T.
No field-induced magnetization plateau was detected, indicating a continuous spin reorientation due to the strong DM interaction.
The results suggest that karpenkoite is not an ideal system for investigating the field-induced plateau states expected for KAFMs.
Nevertheless, the successful growth of single crystals may provide a way to investigate the ground state of KAFM with strong DM interactions.

\vspace{3mm}

\begin{acknowledgment}
We thank Y. Haraguchi for valuable discussions.
This work was supported by JSPS KAKENHI (Nos. JP23H04861 and JP24K06944).
The XRD and pulse-field experiments were performed as joint research at the Institute for Solid State Physics, UTokyo (Project Nos. 202410-MCBXG-0002 and 202504-HMBXX-0042).
We also used the Rigaku XtaLAB Synergy-R at the Molecular Structure Analysis Section, Shizuoka Instrumental Analysis Center, Shizuoka University.
\end{acknowledgment}


\begin{thebibliography}{99}
\bibitem{comment1} $\theta_\mathrm{W} = -17.4$ K was estimated from a Curie–Weiss fit in the 150–300 K range, where the effective magnetic moment of Co$^{2+}$ also varies.
\end{thebibliography}
\end{document}